# The Canadian VirusSeq Data Portal & Duotang: open resources for SARS-CoV-2 viral sequences and genomic epidemiology

## 1.1 Author names

*Erin E. Gill[1], Baofeng Jia[1], Carmen Lia Murall[2,3], Raphaël Poujol[4], Muhammad Zohaib Anwar[5], Nithu Sara John[5], Justin Richardsson[6], Ashley Hobb[7], Abayomi S. Olabode[8], Alexandru Lepsa[6], Ana T. Duggan[3], Andrea D. Tyler[3], Arnaud N'Guessan[9,10], Atul Kachru[6], Brandon Chan[6], Catherine Yoshida[3], Christina K. Yung[11,6], David Bujold[12,13], Dusan Andric[6], Edmund Su[6], Emma J. Griffiths[5], Gary Van Domselaar[3], Gordon W. Jolly[3], Heather K.E. Ward[7], Henrich Feher[6], Jared Baker[6], Jared T. Simpson[6], Jaser Uddin[6], Jiannis Ragoussis[10], Jon Eubank[6], Jörg H. Fritz[14], José Héctor Gálvez[6], Karen Fang[7], Kim Cullion[6], Leonardo Rivera[6], Linda Xiang[6], Matthew A. Croxen[15,16,17,18], Mitchell Shiell[6], Natalie Prystajecky[19,20], Pierre-Olivier Quirion[13], Rosita Bajari[6], Samantha Rich[6], Samira Mubareka[21], Sandrine Moreira[22], Scott Cain[6], Steven G. Sutcliffe[23], Susanne A. Kraemer[24,10], Yann Joly[25], Yelizar Alturmessov[6], CPHLN consortium\*, CanCOGeN consortium\*, VirusSeq Data Portal Academic and Health network\*, Marc Fiume[7], Terrance P. Snutch[26], Cindy Bell[27], Catalina Lopez-Correa[27], Julie G. Hussin[9,28,29], Jeffrey B. Joy[30,31,32], Caroline Colijn[33], Paul M.K. Gordon[34], William W.L. Hsiao[5], Art F.Y. Poon[8], Natalie C. Knox[3], Mélanie Courtot[6,35], Lincoln Stein[6], Sarah P. Otto[36], Guillaume Bourque[12,13], B. Jesse Shapiro[23], Fiona S.L. Brinkman[1,†]*

## 1.2 Affiliation(s)

1. Department of Molecular Biology and Biochemistry, Simon Fraser University, Burnaby, BC, Canada
2. Department of Microbiology and Immunology, McGill University, Montreal, QC, Canada
3. National Microbiology Laboratory, Public Health Agency of Canada, Winnipeg, MB, Canada
4. Research Centre, Montréal Heart Institute, Montréal, QC, Canada
5. Centre for Infectious Disease Genomics and One Health, Faculty of Health Sciences, Simon Fraser University, Burnaby, BC, Canada
6. Ontario Institute for Cancer Research, Toronto, ON, Canada
7. DNAstack, Toronto, ON, Canada
8. Department of Pathology and Laboratory Medicine, Western University, ON Canada
9. Département de Biochimie et Médecine Moléculaire, Université de Montréal, Montreal, QC, Canada
10. McGill Genome Centre, McGill University, Montréal, QC, Canada
11. Indoc Systems, Toronto, ON, Canada
12. Department of Human Genetics, McGill University, Montréal, QC, Canada
13. Canadian Centre for Computational Genomics, Montréal, QC, Canada
14. Department of Microbiology and Immunology, McGill Research Center on Complex Traits (MRCCT), Dahdaleh Institute of Genomic Medicine (DIGM), McGill University, Montréal, QC, Canada
15. Alberta Precision Laboratories, Public Health Laboratory, Edmonton, AB, Canada
16. Department of Laboratory Medicine and Pathology, University of Alberta, Edmonton, AB, Canada
17. Li Ka Shing Institute of Virology, University of Alberta, Edmonton, AB, Canada
18. Women and Children's Health Research Institute, University of Alberta, Edmonton, AB, Canada
19. British Columbia Centre for Disease Control Public Health Laboratory, Vancouver, BC Canada
20. Department of Pathology and Laboratory Medicine, Faculty of Medicine, University of British Columbia, Vancouver, BC, Canada
21. Sunnybrook Research Institute; Department of Laboratory Medicine and Pathobiology, University of Toronto, Toronto, ON, Canada

22. Université de Montréal, Montréal, QC, Canada
23. Department of Microbiology and Immunology, McGill University, Montréal, QC, Canada
24. Aquatic Contaminants Research Division, ECCC, Montréal, QC, Canada
25. Centre of Genomics and Policy, McGill University, Montréal, QC, Canada
26. Michael Smith Laboratories and Djavad Mowafaghian Centre for Brain Health, University of British Columbia, Vancouver, BC, Canada
27. Genome Canada, 150 Metcalfe Street, Suite 2100, Ottawa, ON, Canada
28. Research Centre, Montréal Heart Institute, Montréal, QC, Canada
29. Mila-Québec AI institute, Montréal, QC, Canada
30. Molecular Epidemiology and Evolutionary Genetics, BC Centre for Excellence in HIV/AIDS, Vancouver, BC, Canada
31. Infectious Diseases, Department of Medicine, University of British Columbia, Vancouver, BC, Canada
32. Bioinformatics Programme, University of British Columbia, Vancouver, BC, Canada
33. Department of Mathematics, Simon Fraser University, Burnaby, BC, Canada
34. Centre for Health Genomics and Informatics, University of Calgary, Calgary, AB, Canada
35. Department of Medical BioPhysics, University of Toronto, ON, Canada
36. Department of Zoology & Biodiversity Research Centre, University of British Columbia, Vancouver BC Canada
*Please see supplementary info for names and affiliations.
†Corresponding author

### 1.3  Corresponding author and email address

Fiona S.L. Brinkman, brinkman@sfu.ca

### 1.4  Keywords

data sharing, mutational analysis, evolutionary biology, open access, viral genomics

### 1.5  Repositories:

VirusSeq Data Portal: https://github.com/cancogen-virus-seq/portal
Duotang: https://github.com/CoVaRR-NET/duotang

## 2. Abstract

The COVID-19 pandemic led to a large global effort to sequence SARS-CoV-2 genomes from patient samples to track viral evolution and inform public health response. Millions of SARS-CoV-2 genome sequences have been deposited in global public repositories. The Canadian COVID-19 Genomics Network (CanCOGeN - VirusSeq), a consortium tasked with coordinating expanded sequencing of SARS-CoV-2 genomes across Canada early in the pandemic, created the Canadian VirusSeq Data Portal, with associated data pipelines and procedures, to support these efforts. The goal of VirusSeq was to allow open access to Canadian SARS-CoV-2 genomic sequences and enhanced, standardized contextual data that were unavailable in other repositories and that meet FAIR standards (Findable, Accessible, Interoperable and Reusable). In addition, the Portal data submission pipeline contains data quality checking procedures and appropriate acknowledgement of data generators that encourages collaboration. From inception to execution, the portal was developed with a conscientious focus on strong data governance principles and practices. Extensive efforts ensured a commitment to Canadian privacy laws, data security standards, and organizational processes. This Portal has been coupled with other resources like Viral AI and was further leveraged by the Coronavirus Variants Rapid Response Network (CoVaRR-Net) to produce a suite of continually updated analytical tools and notebooks. Here we highlight this Portal, including its contextual data not available elsewhere, and the 'Duotang', a web platform that presents key genomic epidemiology and modeling analyses on circulating and emerging SARS-CoV-2 variants in Canada. Duotang presents dynamic changes in variant composition of SARS-CoV-2 in Canada and by province, estimates variant growth, and displays complementary interactive visualizations, with a text overview of the current situation. The VirusSeq Data Portal and Duotang resources, alongside additional analyses and resources computed from the Portal (COVID-MVP, CoVizu), are all open-source and freely available. Together, they provide an updated picture of SARS-CoV-2 evolution to spur scientific discussions, inform public discourse, and support communication with and within public health authorities. They also serve as a framework for other jurisdictions interested in open, collaborative sequence data sharing and analyses.

## 3. N/A, Please See Section 4.

## 4. Significance as a BioResource to the community

Here, we describe two new open source, open-access resources for the scientific community. The first, the VirusSeq Data Portal, is a web portal containing SARS-CoV-2 genomic sequences derived from clinical samples collected from across Canada from the start of the pandemic to the present. Genomic data is accompanied by contextual data such as patient age, gender and sample collection date, plus additional contextual data useful for analyses that are not available elsewhere (like "Purpose of Sequencing", which aids analysis of variant growth, by differentiating cases sequenced due to baseline surveillance, verses outbreaks, border cases, etc.). The second resource, Duotang, is an interactive webpage displaying genomic and epidemiological analyses of the sequences from the Data Portal. Duotang is updated on a weekly basis and presents an overview of the current COVID-19 situation in Canada. Duotang presents a combination of analyses that are (to our knowledge) not otherwise available publicly. The Data Portal is unique in that it provides sequence data that has undergone rigorous quality checks, and the accompanying contextual data is harmonized across Canada's federated healthcare system, and more substantial than that available from other online databases. Beyond Canada, these tools are readily transferable to other jurisdictions interested in genomic, phylogenetic or molecular epidemiological analyses of SARS-CoV-2.

## 5. N/A

## 6. Data summary

A section describing all supporting external data, software or code, including the DOI(s) and/or accession numbers(s), and the associated URL.
VirusSeq Data Portal URL: https://virusseq-dataportal.ca/
VirusSeq Data Portal GitHub: https://github.com/virusseq/
Duotang GitHub: https://github.com/CoVaRR-NET/duotang
Duotang URL: https://covarr-net.github.io/duotang/duotang.html
DataHarmonizer GitHub: https://github.com/cidgoh/DataHarmonizer

# 7. Introduction

As the COVID-19 pandemic began to unfold, different countries and jurisdictions developed their own SARS-CoV-2 whole genome sequencing capacity, quality standards, analysis pipelines, and methods to release sequence data and accompanying contextual data for public use.

In Canada, these activities were spearheaded by federal funding to Genome Canada, a not-for-profit organization. Genome Canada supported the creation of the Canadian COVID-19 Genomics Network (CanCOGeN), a Canadian consortium for sequencing SARS-CoV-2 (VirusSeq) and its human host (HostSeq). This pan-Canadian network was composed of academics, national, provincial, and territorial public health labs, hospitals, research institutes, and industry. The VirusSeq component of the consortium aimed to track viral spread and evolution through the sequencing and analysis of viral genomes collected and sequenced in Canada. It relied on participation of the Canadian public to obtain the samples and data necessary for its work. The initial goal was to sequence up to 150,000 viral genomes, which was readily surpassed. As of March 2024, the network had sequenced and shared over 550,000 SARS-CoV-2 genomes.  These data are increasingly used for integrated analysis of genome evolution across Canada's jurisdictions, for example, as presented in the Duotang to identify emerging variants, estimate their growth advantage, and relate genomic trends to those observed in publicly available case statistics.

## 7.1   Sequencing and Data Sharing

The Canadian public health system is decentralized and federated, where each province and territory has its own unique healthcare system and is a participant in the Canadian Public Health Laboratory Network (CPHLN). As part of the CanCOGeN consortium, the members of the CPHLN, as well as the Public Health Agency of Canada's National Microbiology Laboratory (PHAC-NML) and regional health units and hospitals, strengthened their sequencing capacity and expertise. In parallel, others in the consortium worked on developing data sharing agreements and engaging the community in the adoption of effective, uniform and secure data sharing practices. For example, at the outset of the pandemic, each province and territory (and sometimes regional health authorities within provinces) used its own case report form to collect contextual data with its viral samples (e.g., patient age bracket, gender, etc.)(1). Once samples have been sequenced by regional public health laboratories (CPHLN branches), their academic and other partners or PHAC-NML, viral genomes and accompanying data were then shared with PHAC-NML and publicly accessible databases.

The diverse formatting and differences in the types of data collected have complicated cross-jurisdictional sequence comparisons. Analyses were also hampered by the speed with which sequence data was shared to public repositories such as GISAID(2–4), which varied greatly from one region to the next and earlier in the pandemic averaged nearly 3 months across Canada (through May 2021; (5)). Support from NML through provincial partners, including deployed Genomics Liaison Technical Officers (GLTOs), coordinated reagent procurement, and innovation funding, enabled accelerated sequencing and data movement to achieve and maintain a median time from collection to submission of less than 21 days since December 2021.  While GISAID has been a powerful resource for facilitating global data integration, it is not fully open and has restrictive user access and terms of use. In the Canadian context, this means that downloading sequences from GISAID, from adjacent provinces or territories, analyzing, and publicly presenting them side-by-side would have violated the terms of the organization, which require a lengthy process of pre-approval by the different data providers for use of unpublished data. Given some controversy that has surrounded GISAID, involving its management, transparency, and access concerns (6,7), sustaining an alternative platform for sequence data sharing has become a priority.

The VirusSeq Data Portal is an open-access, open-source web-based resource that contains SARS-CoV-2 genome sequence data from across Canada that have been collected from the start of the pandemic to the present. The sequence data are available for download to anybody without the necessity of creating an account. Critically, the sequence data are linked with a set of contextual data such as region of origin, date of collection, reason for sequencing, patient age bracket and gender. These contextual data are standardized so that they appear in the same format for each record, and sequence data undergo rigorous quality checks.

The VirusSeq Data Portal serves an integral purpose in that it allows data from different regions to be analyzed jointly in real time. In certain cases it also provides contextual data that are absent from other large public databases, permitting analyses not possible elsewhere. For example a more robust evaluation of variant growth can factor in "Purpose of Sequencing" (whether the case was due to baseline surveillance, targeted surveillance, surveillance of international border crossing, Cluster/Outbreak investigation, etc.)). The use of a DataHarmonizer tool(8) and uploader for FASTA files facilitates more standardized data curation and aids the deposition process for data producers, so contextual data fields collected using different forms can be integrated into a single searchable database. The DataHarmonizer also simplifies sequence submission to other public databases should the data provider choose to do so.

This Portal, with its focus on enriched contextual data and genomic epidemiology analyses, compliments other resources developed like the European COVID-19 Data Portal (9), which offers multiple sequence analysis workflows and enhanced variant browsing.

## 7.2   A Portal to Variant Discovery

With the advent of SARS-CoV-2 variants of concern (VOCs), Canada launched a funding call for a separate academic entity to tackle the emerging threat posed by new variants: the Coronavirus Variants Rapid Response Network (CoVaRR-Net). The Network was formed in early 2021 with funds from the Canadian Institutes of Health Research (CIHR) with a mandate to "coordinate, facilitate and accelerate rapid research throughout Canada that rapidly answer[s] critical [...] questions regarding variants, such as their increased transmissibility [...]". CoVaRR-Net brings together expertise from a variety of disciplines, including virology, epidemiology, evolutionary biology, sociology, Indigenous relations, and public health to collaborate and share findings with national and international stakeholders. One CoVaRR-Net pillar, the Computational Analysis, Modelling and Evolutionary Outcomes initiative (CAMEO), was set up to focus on computational and mathematical approaches to variant surveillance. CAMEO relies on the publicly available sequencing and contextual data in the Data Portal to track the emergence, introduction, and spread of VOCs within Canada. Modeling tools were developed to estimate how quickly variants might propagate in the near future and which variants were likely to overtake other lineages. These tools continue to be applied on both a regional and national basis, with reference to the global context. Potential variants of Canadian origin are flagged for follow-up, and mutations that could confer traits such as resistance to antivirals or stronger binding to the human ACE-2 receptor are tracked.

As CAMEO formed, the fourth wave was peaking in Canada, driven by Alpha (B.1.1.7) in most jurisdictions and also by Gamma (P.1) in the province of British Columbia. There was a need to search for newly emerging lineages with mutations of note (e.g., Spike:E484K) or mutational jumps. At the time, many of the surveillance activities entailed hands-on analyses, repurposing methods for genome exploration and tree building. Therefore, information in the Data Portal was leveraged to build tools and methods including Pokay (linking mutations to functions, https://github.com/nodrogluap/pokay), CoVizu (phylogenetic representation of lineages, https://filogeneti.ca/covizu/, (10)), COVID-MVP (mutational tracking, https://virusmvp.org/covid-mvp/, (11)), and various in-house mathematical and statistical models. Results of analyses conducted with these tools, such as the spread of mutations within the country and descriptions of novel lineages, were communicated in various media (news articles, social media engagement, public reports) for the general public and public health authorities in Canada.

At the start of the Delta wave in the Fall of 2021, CAMEO teamed up with the PHAC-NML and the Public Health laboratories under CanCOGeN to jointly track the two major Delta lineages growing in Canada: AY.25 and AY.27 (12). The compilation of methods used for this analysis led to the idea of building a pipeline to run in a more automated fashion to improve the speed of tracking and to share the methods for use by any jurisdiction. An R Markdown notebook was chosen to accommodate both modeling and visualization of bioinformatic results with web-viewing, interactivity, and explanatory text. This notebook was dubbed 'Duotang', in reference to Canadian slang for 'workbook' used in schools (a generalized trademark; https://en.wikipedia.org/wiki/Duo-Tang), and launched in early 2022. The accompanying GitHub repository of the network (https://github.com/CoVaRR-NET) houses code for running analyses and data visualizations, as well as recipes for running bioinformatics pipelines, such as building SARS-CoV-2 phylogenetic trees using various methods or tracking the accumulation of mutations over time. An option to password-protect versions of the Duotang notebook is available, in cases where there are privacy concerns or non-public data is being analyzed. The underlying code is readily generalizable to other diseases and can also be used for training purposes for students, public health staff, or new research assistants that joined the network. The tools were used to aid risk assessment for different variants, capitalizing on the federated data to identify variants of interest that, in the Canadian vaccination/immunity context, were undergoing selection in multiple provinces and so unlikely to be growing due to a founder effect or superspreader event (13).

## 8. Methods

We briefly summarize the tools used to create the VirusSeq Data Portal, the data processing steps, and the analyses in Duotang (see https://github.com/CoVaRR-NET/duotang for Duotang code). Figure 1 presents an overview of the data flow from sample collection to the Data Portal then Duotang.

### 8.1 Development of the VirusSeq Data Portal

The Ontario Institute for Cancer Research's Genome Informatics team spearheaded the development and the flexible and scalable deployment of the VirusSeq Data portal using the Overture (14) collection of reusable software microservices, including Ego, Song, Score,

Maestro, and Arranger, and Keycloak (15), a third-party OAuth service. In short, Score manages authorized file transfers to and from an S3 object storage in collaboration with Song, which validates submitted file metadata against an established data model and serves as the single source of truth (SSOT) for the data in the system. After file upload and metadata validation, with the assistance of the event-driven messaging system Kafka, Maestro indexes the information into an Elasticsearch cluster. Finally, Arranger, our GraphQL search API and portal UI generator, consumes the indexed data, making it available to the users through a customized Overture DMS-UI. Together, Ego and Keycloak were used to facilitate the authorization and authentication of users and applications using JSON Web Tokens. In addition to the core components, specialized services (https://github.com/virusseq/) were crafted to cater to the unique requirements of the VirusSeq portal: 1) The Pedigree service was created to fetch and aggregate lineage metadata from Viral AI – a federated network for genomic variant surveillance developed by Canadian biotech company DNAStack (16)(See 8.2), allowing users to search by lineage. 2) Muse was introduced to verify that each viral genome FASTA file was paired with a relevant metadata record, increasing the data quality of the resource. 3) Singularity was developed to manage "data release" bundling and file downloads, enabling users to access large file numbers efficiently on release.

## 8.2 Viral AI Lineage Assignment

Viral AI, developed by DNAstack in response to the COVID-19 pandemic to support access to and facilitate discovery from SARS-CoV-2 data, was used for lineage assignments and additional data queries/tabular visualizations. Assembled SARS-CoV-2 genomes are periodically retrieved from the Data Portal and lineages are assigned using pangolin (Phylogenetic Assignment of Named Global Outbreak LINeages; (17)), in UShER (18) mode to maximize accuracy of assignments. Additionally, since pangolin nomenclature and designations are continuously updated as new variants are sequenced and categorized, upon update to pangolin or its databases, all previously assigned lineages are re-assigned using the latest information. In addition to assigning lineages, the pipeline also produces, for each assembly, the set of sites that differs from the SARS-CoV-2 reference genome. The resulting metadata, variant sites, assemblies, and multifasta are processed through an ingestion pipeline and connected to Viral AI where the data are made publicly available for further analysis and interpretation. Following their ingestion into Viral AI, the lineage metadata are retrieved and added to the VirusSeq Data Portal (Figure 2). Releases of the data are maintained so that historical data with older lineage calls may be retrieved: https://virusseq-dataportal.ca/releases

## 8.3 Data Harmonization and Upload to VirusSeq Data Portal

Data curation and harmonization are crucial steps for the ingestion of data into any database, including the VirusSeq data portal, to ensure that accurate and comprehensive contextual data (sample metadata, epidemiological and methods information) can be associated with SARS-CoV-2 lineages and variants. This process involves collecting, organizing, and managing genomic contextual data from different public health laboratories across Canada to ensure its accuracy, completeness, and reliability. This process requires careful attention to detail, as errors or inconsistencies in data can lead to incorrect conclusions and hinder scientific progress. Furthermore, data sharing permissions differ among public health jurisdictions, and data curators possess an understanding of the ethical, legal, and privacy concerns linked to various datasets and resources (such as databases with restricted access versus those accessible publicly). Within the VirusSeq data portal, a data curator collaborates with both the NML and CPHLN to manage these aspects effectively. These curators have an understanding of the ethical, legal, and privacy concerns linked to various datasets and resources. The data curation and harmonization efforts in the VirusSeq Data Portal are supported by the DataHarmonizer tool (8) developed by The Centre for Infectious Disease Genomics and One Health (See methods 8.3). The tool offers key features like contextual data harmonization, validation and quality assurance, offline functionality and local installation, support for collaboration, adaptability to different pathogens (for instance, antimicrobial-resistant (AMR) bacteria, mpox, influenza), and enhanced usability. This tool simplifies the process of uploading, organizing, and managing contextual data and FASTA sequences and performs validation to ensure that the genomic contextual data adheres to the metadata schema of VirusSeq data portal preventing schema errors. These tools and templates have created a framework that enables analyses used to understand how the virus is transmitted through the population. (For example Duotang and COVID-MVP, described below.)

Additional validation and approval processes are able to identify situations where there are errors, such as duplicates, or the inadvertent inclusion of contextual data that could potentially lead to patient reidentification (such as precise geographical location of sample collection) when combined with other contextual data fields. The robust process of review and approval also empowers data submitters to have control and approval of their data before it eventually is made available to the public. Collectively, the Portal combines robust quality checks, rigorous approval processes, additional contextual data, and fully open data, complementing other international resources.

## 8.4 Duotang website for genomic epidemiology analyses and mathematical modeling

Duotang is a collaborative effort involving members of CoVaRR-Net's CAMEO group. It is presented as a RMarkdown notebook summarizing ongoing investigations of the evolution and epidemiology of SARS-CoV-2 VOCs in Canada. On a weekly basis, genomic data and sample metadata are retrieved from the VirusSeq Data Portal and processed via the Duotang workflow. In addition, Canadian case count data is retrieved from the Canadian HealthInfo Database (19). For phylogenetic reconstruction, the sequences are first aligned against the reference genome (Genbank accession NC_045512) with minimap2 (Ver. 2.17;(20)) using a script adapted from CoVizu (10), and subsampled to 1) up to 10,000 samples for non-XBB lineages and 2) all samples from lineages that descend from the XBB recombinant, for which a separate tree is constructed. This sampling is performed three times. A maximum likelihood tree is then reconstructed from each subsample using the COVID-19 release of IQ-TREE (Ver. 2.2.0; (21,22)) assuming a general time reversible model (GTR with unequal rates and unequal base frequencies) and parameters (-me 0.05 -nt 8 -ninit 10 -n 8). TreeTime (23) is used to reconstruct a time-scaled tree under a strict molecular clock model (the rate is estimated by TreeTime). The resulting trees are converted into an interactive web element within the RMarkdown HTML using ggfree (Ver. 0.2; (24)) and r2d3 (Ver. 0.2.6; (25)). Root-to-tip (RTT) regressions are performed by rooting the tree on the reference genome and then fitting robust linear models to the divergence and sample dates within each major clade. Interactive plots of RTT results and streamgraphs of variant frequencies are generated with r2d3 and custom JavaScript. For other plots (i.e., lineage frequency and selection), the metadata are ingested into R (Ver. 4.1.3) and transformed using dplyr (Ver. 1.1.4) and plotted using ggplot2 (Ver. 3.4.4). For interactivity of these plots within the notebook, Plotly R (26) is used. The RMarkdown notebook is compiled into a HTML page and hosted on Github Pages and CloudFlare Pages for the public. For private Duotang pages that can contain additional non-public data, the web page is password encrypted with Python (Ver 3.9) and the Python library PyCryptoDome (Ver. 3.16) using SHA256 before being published via Github Pages.

## 8.5 Estimating SARS-CoV-2 Variant of Interest (VOI) substitution rates

Substitution rates are obtained from the maximum likelihood tree made using IQ-TREE (see 8.4) and root-to-tip regression performed as described above (see 8.4), without forcing the intercept to zero (similar results are seen when forcing the intercept). For the estimation of each VOI's substitution rates over time, all sequences of that VOI present in the tree are used. While this ignores pseudo-replication among the samples due to relatedness, the estimated slope is robust given the large sample sizes that capture multiple mutational events (see (27)). Standard error (SE) bars are calculated based on the three independent sub-samples to reduce the influence of closely related viral samples. For comparison, a global rate estimate is obtained from a regression over time of all the sequences present in the tree, ignoring variant classifications.

## 8.6 Estimating Selection Coefficients of Variants of Interests.

Selection coefficients provide a measure of the rate of spread of a lineage, relative to other lineages. To estimate selection, we use standard likelihood techniques. In brief, sublineages of current interest are specified (e.g., XBB.1.5, EG.5.1, HV.1), and counts by day tracked over time. If selection were constant over time, the frequency of sub-type *i* at time *t*, measured in days, would be expected to rise according to:

$$p_i(t) = \frac{p_i(0) exp(s_i t)}{\Sigma_j p_j(0) exp(s_j t)} \quad (1)$$

where $s_i$ is the selection coefficient favoring sub-type *i* per day and the index *j* ranges over the circulating subtypes.

To provide a consistent measure of selection, Duotang uses a specific variant as a reference over a window of time (e.g., four months), chosen to be the variant that predominated early in this window. The selection coefficient of this variant is set to 0, $s_j$ = 0. Given a reference lineage to which a lineage is being compared, the same logic behind equation (1) gives

$$ln\left(\frac{p_i(t)}{p_{ref}(t)}\right) = ln\left(\frac{p_i(0)}{p_{ref}(0)}\right) + s_i t. \quad (2)$$

Thus, plotting the above over time (a logit plot), selection generates a linear rise over time, whose slope is the strength of selection $s_i$ For example, a selection coefficient of $s_i$ = 0.1 implies that subtype *i* is expected to rise from $p(0)$ = 10% to $p(t)$ = 90% in frequency in 44 days (that is, in 4.4 / $s_i$ days), when considering only the subtype and reference. By contrast, other tools estimate selection on one lineage relative to all remaining lineages, as in equation (1) (e.g., the default setting in CoV-Spectrum (28)), but genetic changes in other lineages then cause selection to appear to vary over time.

At any given time *t*, the probability of observing $n_i$. sequences of sublineage *i* and $n_{ref}$. sequences of the reference sublineage is binomially distributed, given the total number of sequences from that day and the frequency of each $p_i(t)$. Consequently, the likelihood of seeing the observed lineage frequencies over all times *t* and over both sublineages *j* is proportional to:

$$L = \prod_t \prod_j p_j(t)^{n_j(t)} \quad (3)$$

assuming random sampling of cases and the evolutionary model (1). The BBMLE (Ver. 1.0.25; (29)) package in R was used to maximize the likelihood of the observed data to estimate the unknown parameters $s_i$ and $p_i(0)$ relative to the reference (using the default optimization method, optim). For each selection coefficient, 95% confidence intervals are estimated by computing the inverse of the Hessian matrix obtained from the maximum likelihood estimation. Then 95% confidence bands are obtained by randomly drawing 10,000 sets of parameters ($p_i$ and $s_i$ for each sub-type) using RandomFromHessianOrMCMC (30), assuming a multi-normal distribution around the maximum likelihood point (estimated from the Hessian matrix). Graphs illustrating the rise in frequency of a variant over time are shown (see 9.3.2). Extensions comparing a set of variants to the reference is performed analogously, using a multinomial distribution and extending equation (3) for *j*>2.

To estimate case count trends by variant, reported cases obtained from Health Infobase Canada (weekly) or using provincial data sources (daily) where available (Alberta (31) and Quebec (32)) are normalized to cases per 100,000 individuals using Statistics Canada's population estimates (Sept. 27; (33)). The last case data point for each region is then removed, as these data are still being gathered and are thus underestimated. The normalized cases over time ($n(t)$) are then log transformed and fitted to a smooth spline with a lambda value of 0.001 using the smooth.spline() function within R.

The previously discussed methods allow us to estimate the proportion of a lineage over time (see equation for $p_i(t)$ above). Multiplying the frequency of a lineage by the total case count gives the inferred number of reported cases that are due to that lineage at each time point:

$$n_i(t) = p_i(t)\, n(t) \quad (4)$$

Finally, once the inferred case counts ($n_i(t)$) of each lineages are calculated, we take the last two days of data from the smooth spline times lineage frequency curve to get $n_i(t)$ and $n_i(t-1)$, from which the current exponential growth rate of that lineage ($r_i$) is estimated as

$$r_i = ln\left(\frac{n_i(t)}{n_i(t-1)}\right) \quad (5)$$

The resulting estimates are then plotted using ggplot2.

## 9. Results

### 9.1 The VirusSeq Data Portal: an Open-Access SARS-CoV-2 Sequence Repository

In response to the COVID-19 pandemic, the initial version of the VirusSeq Data Portal database schema (https://virusseq-dataportal.ca/) was developed and deployed in a rapid four-week timeframe. (See Methods 8.1). OICR's repurposable open-source data management software, Overture, played a crucial role in streamlining the development process for the project. The modular architecture was particularly beneficial, allowing the system to scale in response to unexpected data submission surges. There were initial delays in receipt of data, as important data sharing agreements and procedures were developed that supported federated public health-academic collaboration. Ethics review and surveys of the public were also completed, to help inform appropriate data sharing policies (34–36). As of March 2024, the data portal houses >550,000 SARS-CoV-2 viral genomes and their associated contextual data (metadata) collected across Canadian provinces, with more being added on a weekly basis. Contextual data include anonymized information regarding the patient from which samples were collected (e.g., age bracket, gender), geographic and temporal information (e.g., province/territory of origin, sample collection date), epidemiological information (e.g., purpose of sampling , purpose of sequencing), sequencing and bioinformatics information (e.g., sequencing protocol, dehosting method), viral lineage information and a range of additional data fields (see Supplementary Information for details). The portal satisfies the needs of three primary user groups: data administrators, submitters, and public users. Public users can access the data via an API or the user-friendly web portal, which facilitates data filtering using a diverse range of metadata fields (Figure 3). Users can then download the filtered metadata alone or alongside the associated FASTA files. Data providers with authorization to submit have a dedicated graphical user interface to upload genomic data and associated metadata. Lastly, data administrators can access back-end services that control data access, enable maintenance and data configuration, and manage user authorization and authentication. The development of the VirusSeq Data Portal is a testament to the power of collaborative efforts, modular design, and the reusability of software tools in bioinformatics. This design, and in turn the portal has already been reused in additional projects to track pathogen genomics (e.g., (37)).

Lineage assignments, in combination with sample metadata and assemblies, are also imported into Viral AI where they are made available over GA4GH (Global Alliance for Genomics and Health) standard interfaces. In addition to assigning lineages, the pipeline also produces, for each assembly, a set of sites that differs from the SARS-CoV-2 reference genome (see supplementary methods). The resulting metadata, variant sites, assemblies, and multiFASTA files are processed through an ingestion pipeline and connected to Viral AI, with the data made publicly available for further analysis and interpretation (https://viral.ai/collections/virusseq/overview). Finally, metadata are retrieved and added to the VirusSeq Data Portal.

## 9.2  Duotang: A Multi-Faceted Resource for Tracking Variants

Duotang is enabled by the steady stream of open-access, harmonized data deposited into the VirusSeq Data Portal (see section 9.2 above). This deposition is powered by data generators across Canada who are willing to openly share anonymous data. The only criteria for use is to acknowledge the contribution of VirusSeq and its partners in any publication (see acknowledgements & consortium and network author information in supplementary material).

The various analyses and tools that Duotang offers include a manually curated text summary of the current situation, reviewed by multiple CAMEO members, plots of major sublineages that have been circulating in the country within the last 120 days, selection estimates on groups of lineages, case count trends by variant, selection estimates and 95% confidence intervals for the fastest growing pangolin lineages (see sections 8.1 and 9.2 above), focusing on variants with more than 10 sequences in Canada over the past 4 months, and highlighting those undergoing positive selection in multiple provinces. It also shows phylogenetic trees, root-to-tip analyses, and molecular clock estimates. The majority of these visualizations are interactive, allowing the user to select distinct lineages/sublineages, Pango groups, or plot types. Furthermore, Duotang is updated weekly, as data is being released in the VirusSeq Data Portal. The implementation of these analyses and their visualizations are described in Methods 8.4-8.5. Here, selected features of Duotang are briefly described. Please refer to Figure 4 for an illustrated introduction to Duotang.

### 9.2.1 Current SARS-CoV-2 situation and summarizing variants

For each update, a snapshot summary of the current SARS-CoV-2 landscape within Canada is prepared by expert reviewers. Then the most frequent 50 sublineages of SARS-CoV-2 present in Canada are plotted, categorized by week. The frequencies are expressed both as an absolute sequence case count as well as a relative count.

### 9.2.2 Estimating selection on variant of interest and integrating case counts

Duotang also provides a visualization of the estimated rate of spread of lineages, specifically, visualizing if new or emerging lineage(s) have selective advantages, and by how much, relative to a previous predominant lineage over the last four months. Thus, we can determine if new or emerging lineage(s) have selective advantage(s), and by how much. Selection coefficient estimates have become standard metrics of new variant fitness and an early indicator of potential impact, which are now incorporated in global technical reports. Tracking selection coefficients can show accelerating or decelerating spread. Such changes in selection have been observed following widespread reductions in transmission, such as following a "circuit breaker" in British Columbia Canada on March 30 2021 aimed at halting the spread of Alpha and Gamma(38). Outbreaks or other localized transmission can also lead to a rise in frequency of a particular variant, but such a rise is restricted in time and place. Duotang, thus, highlights those variants that are observed to rise in frequency across time – and across multiple provinces – allowing variants with a true selective advantage to be distinguished from variants that rise in frequency by chance. Moreover, by integrating this selection coefficient with clinical PCR-based case data, we can estimate the proportion of cases that can be attributed to a specific variant and the current growth rate of each variant (Figure 4).

### 9.2.3 Phylogenetic Trees, Root-to-Tip Analyses, and Molecular Clock Estimates

Duotang also presents interactive phylogenetic based analyses, including time and diversity trees and molecular clock estimates (Figure 4). The interactivity of these analyses allows the user to quickly identify samples that are outliers but may hold clinical significance, (e.g., identification of highly divergent sequences or lineages that have re-emerged after a long absence, both of which may indicate passage through a chronic infection or another host species). Furthermore, the slope of the root-to-tip plots over time provide an estimate of the substitution rate. A lineage with a steeper positive slope than average for SARS-CoV-2 is accumulating mutations at a faster pace, while a lineage that exhibits a jump (a shift up in intercept but not slope) has accumulated more than expected numbers of mutations in a transient period of time (similar to Alpha when it first appeared in the UK).

## 9.3 Additional Resources

In addition to Duotang, the open-source data within the VirusSeq Data Portal has powered other tools. One such example is CoVizu that illustrates the mutational connections among sampled lineages (https://virusseq-dataportal.ca/visualization; https://github.com/PoonLab/CoVizu). Another is COVID-MVP (https://virusmvp.org/covid-mvp/) that connects mutations with their functional impact. It is an interactive heatmap-based visualization app that allows users to explore the mutational profile of a set of genomes, including the underlying literature on the functional impact linked with each mutation. Functional annotation is a continuous effort that is primarily performed by curators working on the CAMEO initiative, however, the scientific community can also provide input by using the issue tracking system of the GitHub repository (https://github.com/nodrogluap/pokay), where the annotations are maintained. Comprehensive details on COVID-MVP features including the visualization, backend genomics workflow, functional annotations, and deployment of the framework to individual systems or cloud-based HPCs are described in (11).

### 9.4 Ethical Considerations for the VirusSeq Data Portal

Article 27 of the 1948 Universal Declaration of Human Rights guarantees the rights of every individual in the world "to share in scientific advancement and its benefits" (39). The sharing of genomic and health-related data is of key importance for pandemic monitoring and prevention for the benefit of human health. The VirusSeq Ethics and Governance Working Group drafted a policy memorandum along with a scientific manuscript (34) to clarify the Canadian ethical and legal framework applicable to the privacy implications of sharing viral genetic sequences and engage the Canadian public health and research community and other stakeholders. The conclusion of this work was that the sharing of viral genetic sequences that have been carefully de-hosted of any human-like or non-viral sequences through the application of robust technical standards is unlikely to create any privacy risk for the host of the virus. Adding to this information a minimal amount of non-identifying metadata (e.g., age binned by category, gender, instrument used for the collection, etc.) does not change this general conclusion. By providing clear and concise information on the legal and ethical aspects of data sharing, the Working Group aimed to have a positive impact on Canadian practice.

To ensure the VirusSeq Data Portal would be compliant with the highest ethical and legal standards, we created a data security committee responsible for assessing the security, integrity, privacy, and ethical compliance of the Portal in an ongoing and forward-looking manner. The Committee participated in all team meetings, allowing any ethical or legal concern to be quickly addressed by the Portal team. The presence of a data security committee aimed to increase the confidence of stakeholders contributing data to the Portal.

## 10. Figures and tables

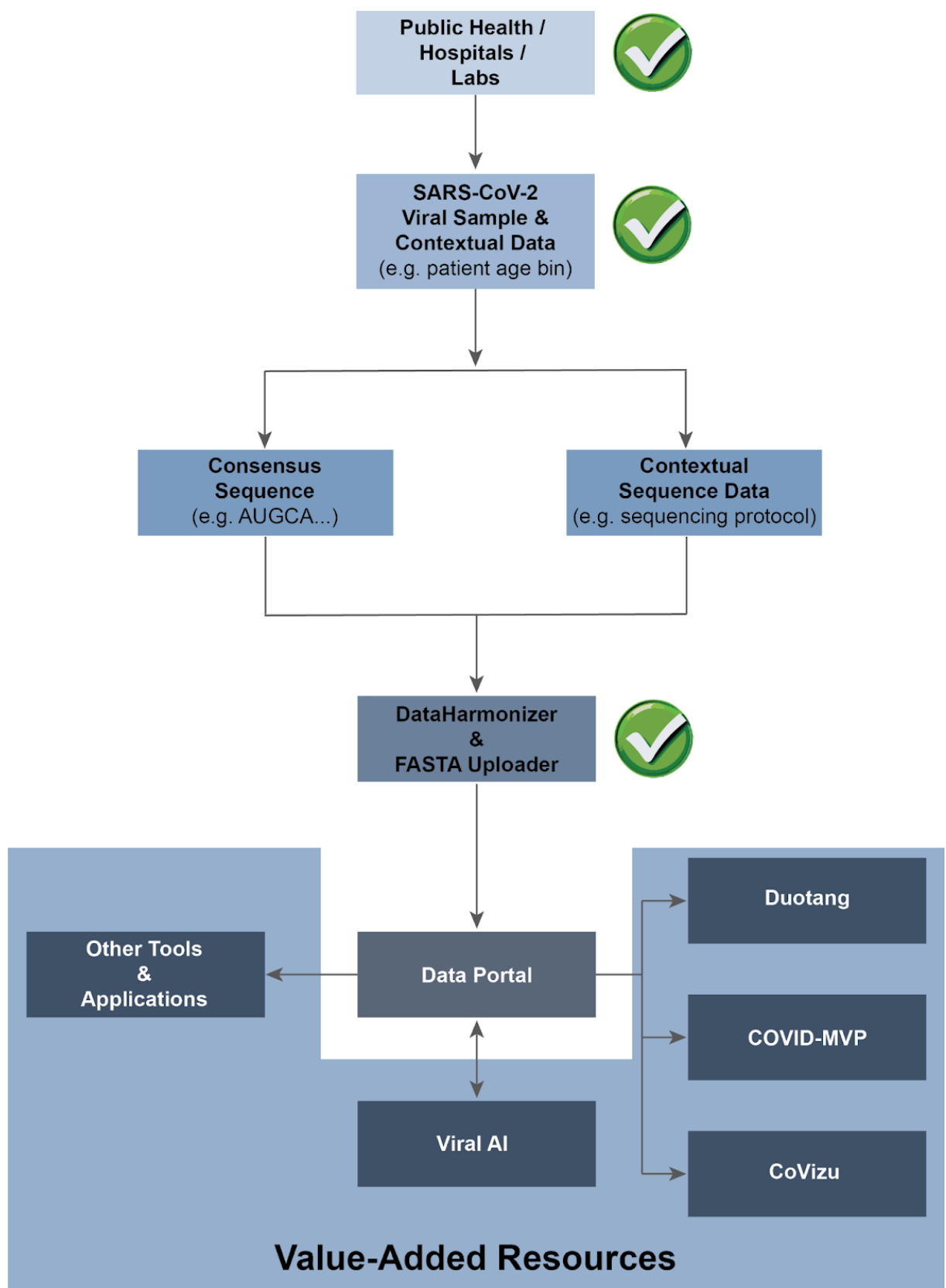

*Figure 1: Data Flow Overview.* SARS-CoV-2 viral samples and contextual data are collected by Public Health, hospitals and labs. Some samples are selected for sequencing based on national and regional priorities. After sequencing, regional Public Health authorities select

samples that will be shared publicly. Sequence and contextual data from these samples are uploaded to the Data Portal using the DataHarmonizer and FASTA Uploader. From the Portal, the public can view or download contextual and sequence data from across the country. Data can also be accessed for value-added resources, such as Duotang, ViralAI, COVID-MVP and CoVizu via the API. Green check marks on the figure indicate points in the data flow where human Quality Assurance / Quality Control (QA/QC) and/or ethics oversight take place.

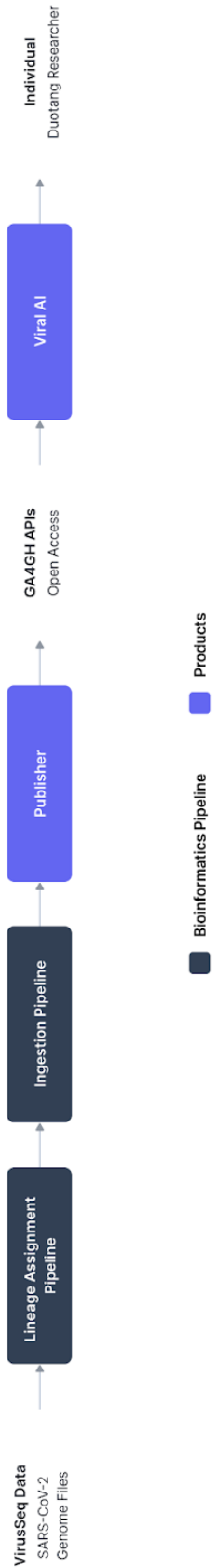

***Figure 2: Overview of the data flow from VirusSeq Data Portal to Duotang.*** Genomics and epidemiological data from VirusSeq is first processed by DNAStack's Viral AI workflow. The result of this is a dataset containing SARS-CoV-2 lineage information. Duotang then retrieves this lineage information using Viral AI's Global Alliance for Genomics and Health (*GA4GH*) compliant APIs for further analyses.

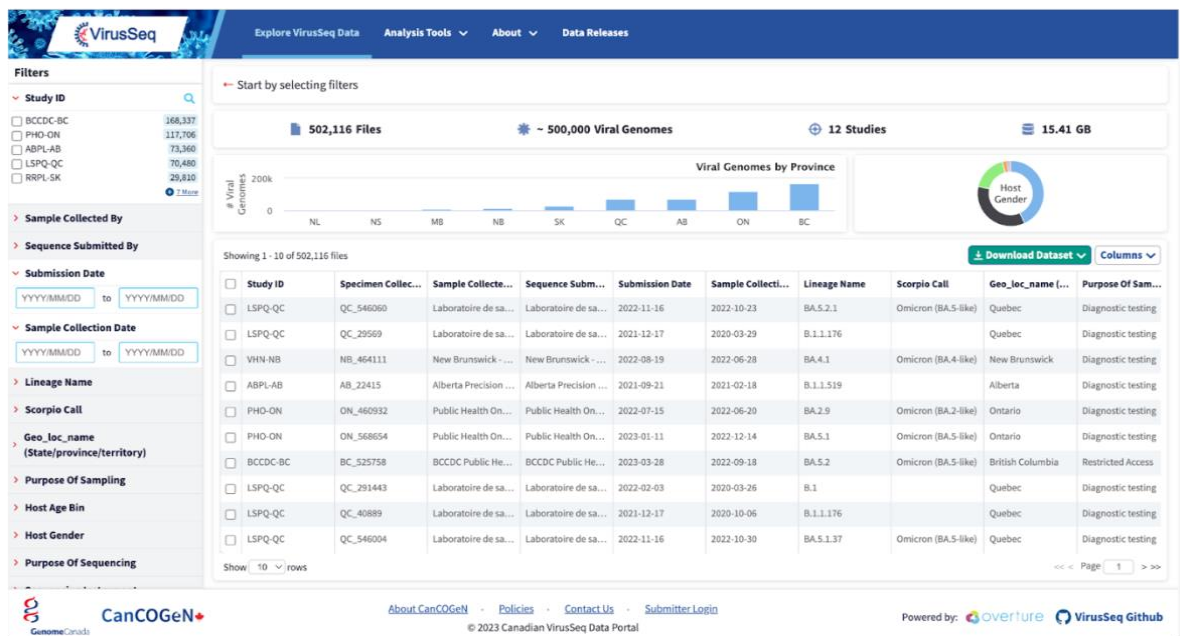

*Figure 3. Overview of the VirusSeq Data Portal Explore page.* The VirusSeq Data Portal allows users to browse samples stored via a web interface or API. Within the interface, the user is able to see the samples that are available, their metadata, and perform filters and queries to identify samples of interest that can then be downloaded for analysis.

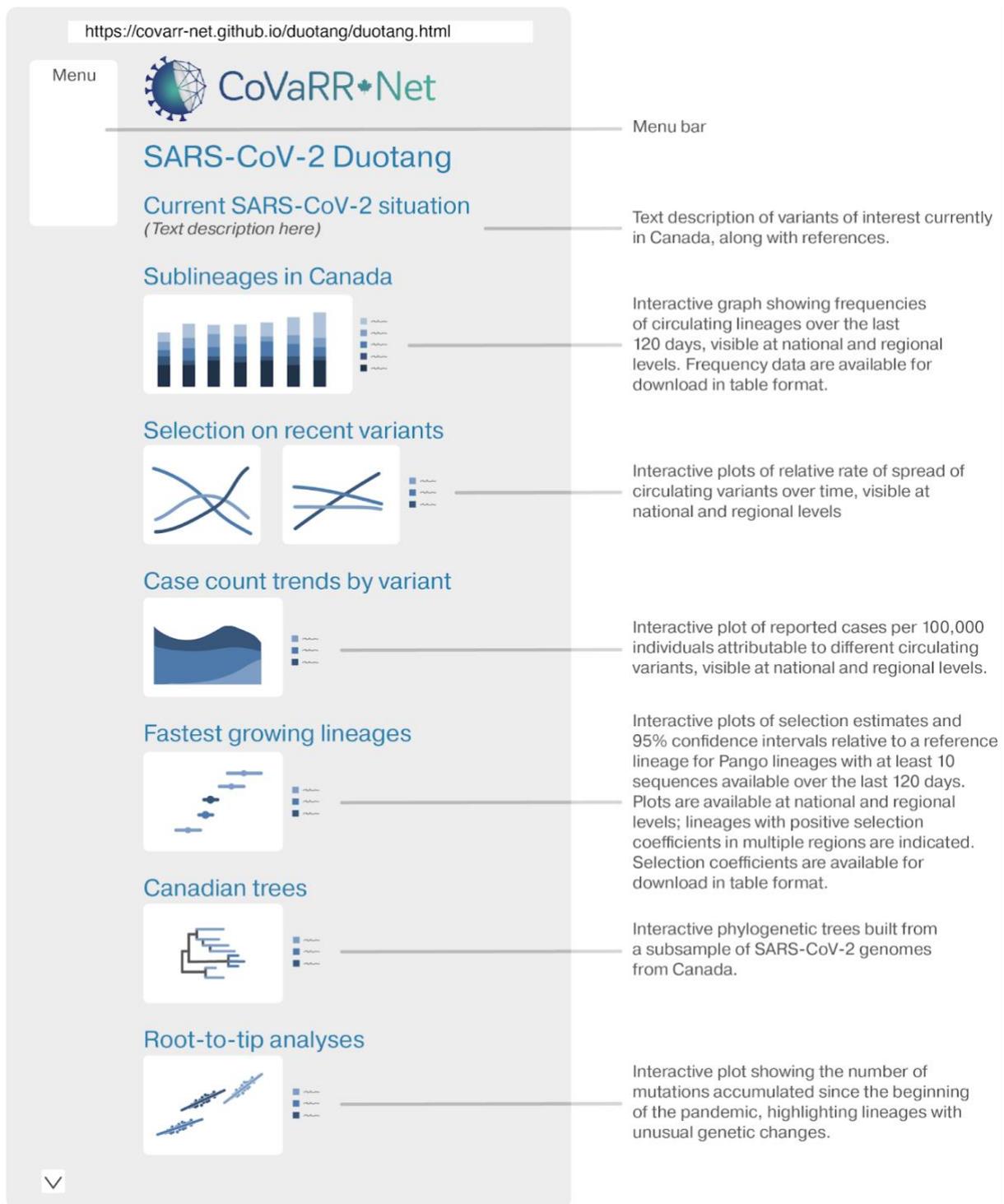

***Figure 4: Duotang Webpage Overview.*** Duotang contains many different interactive plots for the user to explore SARS-CoV-2 genomic epidemiology in Canada. All sections of the page are easily accessible via the menu bar, which is located on the left-hand side of the page. The first section of the page gives a text description of current variants of interest in the country. Several of the plots are highlighted in the figure above, which include different visualizations of selection on variants, phylogenetic trees and root-to-tip analyses of these trees to detect unusual genetic changes. In addition to what is shown here, the user can also view plots of the growth advantage of single lineages relative to a reference lineage (on both national and regional levels), visualize the mutational composition of actively circulating lineages via an embedded frame of COVID-MVP, review the changing proportions of different variants over time (back to April 2020, on both national and regional levels), examine molecular clock estimates for different VOIs, and utilize a searchable table to view the ancestors and description of any Pango lineage. For more details, please visit: https://covarr-net.github.io/duotang/duotang.html

## 11. Discussion

### 11.1 Benefits of Open Data

The VirusSeq Data Portal and Duotang are open-source and open-access resources that allow users to discover and download SARS-CoV-2 genomic data and explore mathematical models and genomic epidemiology analyses performed on the aforementioned data. The VirusSeq Data Portal simplifies Canadian SARS-CoV-2 genomic data discoverability and retrieval for users by providing a single, central repository where sequences are available from a federated, provincial health care system. Duotang allows users to see diverse types of analyses on an interactive webpage, preventing individuals from having to switch between various sites to find what they're looking for. While developed for Canada, the tools may be readily applied to other jurisdictions. The benefits of such open data have been described elsewhere (38) and can be summarized as 1) increasing scientific collaboration and innovation, 2) supporting research and potential for increased analyses, 3) informing public policy and decision-making, 4) allowing the public to be better informed, and 5) enabling scientific reproducibility. Indeed, the Portal has permitted analyses (12,40) that would not otherwise have been possible. The Portal has also allowed the development of Duotang, which is used as a resource by PHAC-NML, the CPHLN, and researchers who present modeling updates to government officials, thus informing the development of public policy and accelerating innovation. Industry has also used it to help inform vaccine development, and academics have used it to inform research priorities. None of this would have been possible without a commitment to open data by a range of Public Health and hospital workers and academic researchers. The development of the VirusSeq Data Portal is overall a testament to the power of collaborative efforts, modular design, and the reusability of software tools in bioinformatics.

### 11.2 Challenges and Solutions

In Canada, as in other countries, data silos existed at the beginning of the pandemic. This was largely due to the aforementioned federated nature of the Canadian provincial healthcare system. Once sequencing capacity was built and different regions began to produce their own data, regions faced uncertainty regarding the legal and ethical requirements applicable to data sharing, how to prevent the re-identification of patients, as well as about the kind of rigorous quality control steps that would ensure only high-quality data were shared. These challenges were overcome through careful consultation and agreement on clear policies.

The existing contextual data that is available from the Portal is sufficient to perform many useful analyses (such as those in Duotang), but additional (and linked) contextual data could allow us to determine (for example) how effective our vaccines are against current variants in a Canadian context, which variants spread more easily in certain contexts (such as among children), or which variants are more severe. Linking sequence data with epidemiological, clinical and immunization data would allow us to better utilize and leverage the sequence and minimal contextual data we already have (41). This is not a simple undertaking in Canada, as these types of data are often collected by different parts of the health care system. For privacy reasons, linked data would have to be made available in a way that would afford reasonable privacy protection from re-identifiability to patients and research participants. A first study of Canadian participants' preference on the matter indicates a general willingness to share their data in a pandemic context (35). Another option would be to ensure that any linked, potentially identifiable data would be made available solely to vetted researchers.

### 11.3 The Future of Duotang and the Data Portal

Collectively, the Portal and associated resources aim to provide open SARS-CoV-2 sequence data and analyses that have undergone rigorous quality checks, with harmonized contextual data from across the country that is more substantial than that available from other online databases. With the focus shifting back to other viruses of concern, Duotang is being adapted into a framework that is capable of performing similar analyses for other emerging pathogens. Currently, an extension of Duotang, which houses data for influenza virus and mpox virus, is being reviewed internally. Furthermore, with the emergence of pathogen surveillance via environmental methods (e.g., wastewater sampling), Duotang can be further extended with metagenomics data in mind. Similarly, the Data Portal can be re-tooled in the future to enable storage and comparison of wastewater data. These enhancements will enable Canadians and others internationally to access and analyze a wide variety of emerging pathogen surveillance data, develop tools to improve analyses using these data, and provide templates for additional bona fide services for tracking infectious diseases of concern.

## 12. Author statements

### 12.1 Author contributions

**Abayomi S. Olabode:** Methodology, Software, Validation, Writing-reviewing and editing, Visualization **Alexandru Lepsa:** Software, Writing-reviewing and editing **Ana T. Duggan:** Methodology, Validation, Resources, Writing-reviewing and editing **Andrea D. Tyler:** Methodology, Validation, Resources, Writing-reviewing and editing **Arnaud N'Guessan:** Software, Resources, Writing-reviewing and editing, Visualization **Art F.Y. Poon:** Conceptualization, Methodology, Software, Formal analysis, Investigation, Resources, Writing-reviewing and editing, Visualization, Supervision **Ashley Hobb:** Software, Writing-original draft, Writing-reviewing and editing **Atul Kachru:** Software, Writing-reviewing and editing, Visualization **B. Jesse Shapiro:** Conceptualization, Methodology, Validation, Writing-reviewing and editing, Supervision, Funding acquisition **Baofeng Jia:** Methodology, Software, Validation, Formal analysis, Writing-original draft, Writing-reviewing and editing, Visualization, Project administration **Brandon Chan:** Software, Writing-reviewing and editing **Carmen Lia Murall:** Conceptualization, Methodology, Software, Validation, Formal analysis, Writing-original draft, Writing-reviewing and editing, Visualization **Catalina Lopez-Correa:** Conceptualization, Writing-reviewing and editing, Supervision, Project administration, Funding acquisition **Caroline Colijn:** Conceptualization, Methodology, Software, Validation, Formal analysis, Writing-original draft, Writing-reviewing and editing, Visualization, Supervision, Funding acquisition **Catherine Yoshida:** Methodology, Validation, Resources, Writing-reviewing and editing **Christina K. Yung:** Conceptualization, Software, Writing-reviewing and editing, Visualization **Cindy Bell:** Conceptualization, Writing-reviewing and editing, Supervision, Project administration, Funding acquisition **David Bujold:** Conceptualization, Software, Writing-reviewing and editing, Visualization, Supervision **Dusan Andric:** Software, Writing-reviewing and editing, Visualization **Edmund Su:** Software, Writing-reviewing and editing, Visualization **Emma J. Griffiths:** Conceptualization, Methodology, Software, Validation, Resources, Writing-reviewing and editing, Supervision, Data curation **Erin E. Gill:** Methodology, Software, Validation, Writing-original draft, Writing-reviewing and editing, Visualization, Supervision, Project administration **Fiona S. L. Brinkman:** Conceptualization, Methodology, Software, Validation, Formal analysis, Writing-original draft, Writing-reviewing and editing, Visualization, Supervision, Project administration, Funding acquisition **Gary Van Domselaar:** Conceptualization, Methodology, Software, Resources, Writing-reviewing and editing, Supervision, Funding acquisition **Gordon W. Jolly:** Conceptualization, Methodology, Resources, Writing-reviewing and editing, Supervision, Funding acquisition **Guillaume Bourque:** Conceptualization, Methodology, Software, Writing-reviewing and editing, Visualization, Supervision, Funding acquisition **Heather K.E. Ward:** Methodology, Software, Writing-original draft, Writing-reviewing and editing, Visualization **Henrich Feher:** Software, Writing-reviewing and editing, Visualization **Jared Baker:** Software, Writing-reviewing and editing, Visualization **Jared T. Simpson:** Software, Validation, Writing-reviewing and editing, Visualization, Supervision, Funding acquisition **Jaser Uddin:** Software, Writing-reviewing and editing, Visualization **Jeffrey B. Joy:** Writing-original draft, Writing-reviewing and editing, Supervision, Funding acquisition **Jiannis Ragoussis:** Methodology, Validation, Resources, Writing-reviewing and editing, Supervision, Funding acquisition **Jon Eubank:** Software, Writing-reviewing and editing, Visualization **Jörg H. Fritz:** Writing-reviewing and editing, Supervision, Funding acquisition **José Hector Galvez:** Methodology, Software, Resources, Writing-reviewing and editing, Visualization **Julie G. Hussin:** Conceptualization, Methodology, Resources, Writing-reviewing and editing, Visualization, Supervision, Funding acquisition **Justin Richardsson:** Software, Writing-reviewing and editing, Visualization **Karen Fang:** Software, Writing-original draft, Writing-reviewing and editing, Visualization **Kim Cullion:** Software, Writing-reviewing and editing, Visualization **Leonardo Rivera:** Software, Writing-reviewing and editing, Visualization **Lincoln D. Stein:** Conceptualization, Methodology, Software, Writing-reviewing and editing, Visualization, Supervision, Funding acquisition **Linda Xiang:** Software, Writing-reviewing and editing, Visualization **Marc Fiume:** Conceptualization, Methodology, Software, Writing-reviewing and editing, Visualization, Supervision, Funding acquisition **Matthew A. Croxen:** Conceptualization, Methodology, Validation, Resources, Writing-reviewing and editing, Supervision, Data curation **Mélanie Courtot:** Conceptualization, Methodology, Software, Validation, Resources, Writing-original draft, Writing-reviewing and editing, Visualization, Supervision, Funding acquisition **Mitchell Shiell:** Software, Writing-original draft, Writing-reviewing and editing, Visualization **Muhammad Zohaib Anwar:** Conceptualization, Methodology, Software, Validation, Formal analysis, Writing-reviewing and editing, Visualization **Natalie C. Knox:** Conceptualization, Methodology, Software, Validation, Resources, Writing-reviewing and editing, Supervision, Data curation **Natalie Prystajecky:** Conceptualization, Methodology, Validation, Resources, Writing-reviewing and editing, Supervision, Data curation **Nithu Sara John:** Conceptualization, Methodology, Software, Validation, Writing-original draft, Writing-reviewing and editing, Data curation **Paul M.K. Gordon:** Conceptualization, Methodology, Software, Validation, Writing-reviewing and editing, Visualization, Supervision, Funding acquisition, Data curation **Pierre-

**Olivier Quirion:** Software, Resources, Writing-reviewing and editing, Supervision **Raphaël Poujol:** Conceptualization, Methodology, Software, Validation, Writing-reviewing and editing, Visualization, Data curation **Rees Kassen:** Conceptualization, Writing-reviewing and editing, Supervision, Funding acquisition **Rosita Bajari:** Software, Writing-reviewing and editing, Visualization **Samantha Rich:** Software, Writing-reviewing and editing, Visualization **Samira Mubareka:** Conceptualization, Resources, Writing-reviewing and editing, Supervision, Funding acquisition **Sandrine Moreira:** Conceptualization, Methodology, Resources, Writing-reviewing and editing, Supervision, Funding acquisition **Sarah P. Otto:** Conceptualization, Methodology, Software, Validation, Formal analysis, Writing-original draft, Writing-reviewing and editing, Visualization, Funding acquisition **Scott Cain:** Conceptualization, Methodology, Software, Validation, Writing-original draft, Writing-reviewing and editing, Visualization, Supervision **Steven G. Sutcliffe:** Validation, Writing-reviewing and editing, Visualization, Project administration **Susanne A. Kraemer:** Software, Validation, Writing-reviewing and editing, Visualization **Terrance P. Snutch:** Conceptualization, Methodology, Resources, Writing-reviewing and editing, Visualization, Supervision, Project administration, Funding acquisition **William W.L. Hsiao:** Conceptualization, Methodology, Software, Writing-reviewing and editing, Visualization, Supervision, Funding acquisition, Data curation **Yann Joly:** Conceptualization, Methodology, Writing-original draft, Writing-reviewing and editing, Supervision, Funding acquisition **Yelizar Alturmessov:** Software, Writing-reviewing and editing, Visualization

## 12.2 Conflicts of interest

J.T.S. receives research funding from Oxford Nanopore Technologies (ONT) and has received travel support to attend and speak at meetings organized by ONT, and is on the Scientific Advisory Board of Day Zero Diagnostics.

## 12.3 Funding information

Funding was gratefully provided by the Canadian COVID-19 Genomics Network (CanCOGeN; Genome Canada grant #E09CMA) and the Coronavirus Variant Rapid Response Network (CoVaRR-Net), supported by Genome Canada, Innovation, Science and Economic Development Canada (ISED) and Canadian Institutes of Health Research (CIHR) (grant #ARR-175622). Work by personnel at the NML was funded by the above CanCOGeN grant and operational funds from the Public Health Agency of Canada. Additional funding was provided via Genome Canada grant #286GET, CIHR grant #174924, IVADO COVID19 Rapid Response grant #CVD19-030, US National Institutes of Health grant #U24CA253529, Canarie, Simon Fraser University (Research Computing Group), the Digital Research Alliance of Canada, and the Ontario Institute for Cancer Research (OICR). Initial funding for SARS-CoV-2 sequencing in Alberta was supported by Genome Alberta COVID-19 Rapid Regional Response Program (ACRRP-5) awarded to MAC and Linda Chui, and then further supported through CanCOGeN and PHAC-NML Canadian COVID-19 Genomics Program (CCGP; rebranded Applied Genomics Innovation for Laboratory Excellence (AGILE)). J.G.H. is a Fonds de Recherche du Québec – Santé (FRQS) Junior 2 research scholar. J.T.S. is supported by OICR through funds provided by the Government of Ontario and the Government of Canada through Genome Canada and Ontario Genomics (#OGI-136 and #OGI-201). M.Z.A is supported by a CIHR Fellowship and a Michael Smith Health Research BC (MSHRBC) trainee Fellowship. F.S.L.B holds a Simon Fraser University Distinguished Professorship.

### 12.4 Ethical approval

N/A

### 12.5 Consent for publication

N/A

### 12.6 Acknowledgements

The results here are in whole, or in part based upon data hosted at the Canadian VirusSeq Data Portal: https://virusseq-dataportal.ca/. We wish to acknowledge the Canadian Public Health Laboratory Network (CPHLN) for the generation and contribution of SARS-CoV-2 sequences and accompanying data to the Portal and the CanCOGeN VirusSeq Consortium for its contribution to the construction of the Portal, creation of data standards and ethical considerations. For complete lists of individuals belonging to the CPHLN and the CanCOGeN VirusSeq consortium, see supplementary information.

# Supplementary Information: The Canadian VirusSeq Data Portal & Duotang: open resources for SARS-CoV-2 viral sequences and genomic epidemiology

## 1. Supplementary Methods

### 1.1 DNAStack Viral AI network for genomic variant surveillance

Viral AI is the world's first federated network for genomic variant surveillance, developed by DNAstack in response to the COVID-19 pandemic to support discovery and access to SARS-CoV-2 data. DNAstack partnered with the CanCOGeN - VirusSeq project to make the VirusSeq data available through Viral AI and to support lineage assignment for tracking variants.

Viral AI introduces a new way to share and analyze genomics, clinical, administrative, and related data, facilitating insights about transmission, severity, diagnostics and vaccine escape. As an alternative to the centralized model, where data is uploaded to a single vendor-managed database, Viral AI adopts a federated architecture to connect, analyze, and share data without moving it. This model enables faster, more efficient, regulatory compliant, and regionally sovereign data management, enabling viral surveillance efforts to be more equitable, scalable, and sustainable (see figure S1).

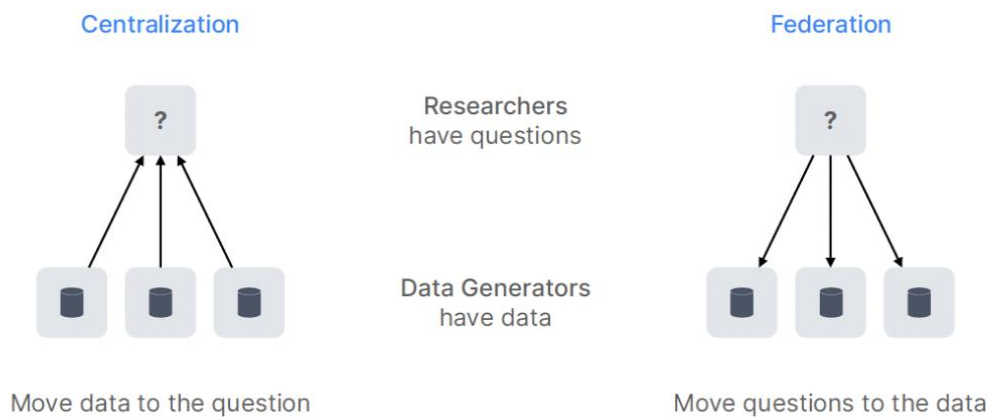

Figure S1: Federation makes it possible to drive discoveries across distributed data without moving it.

Viral AI accelerates science by making data uniformly accessible through a user-friendly graphical interface and powerful programmatic interfaces, integrating data across different sources from around the world alongside VirusSeq, such as NCBI Sequence Read Archive (SRA) and European Center for Disease Prevention and Control, among others. Over one million viral sequences have been added with corresponding assemblies, variant calls, and lineage assignments, all harmonized through an open source bioinformatics pipeline.

Viral AI is powered by a software suite that is compliant with multiple GA4GH standards and facilitates responsible and interoperable genomic and biomedical data sharing including the Data Connect, Data Repository Service, Service Registry, and Service Info standards.

**Publisher** is a data integration and sharing studio that enables data custodians to connect any dataset, from any source, without moving it. Data custodians who contribute data retain administrative control and have transparency into how it's used. DNAstack has connected a number of open source viral genomic data sets using Publisher alongside the VirusSeq data.

**Explorer** is a federated data hub that makes it easier for researchers to find, access, and analyze shared data. With Explorer, researchers can search and perform analyses across a universe of connected datasets through a single user interface. The VirusSeq data is made available in Explorer for researchers to discover, access, and analyze alongside the other connected data sets.

### 1.2  Lineage Assignment Pipeline

An open source bioinformatics pipeline was developed to run lineage assignment on the SARS-CoV-2 genome assemblies obtained from VirusSeq. The resulting lineage assignments, in combination with sample metadata and assemblies, are imported into Viral AI where they are made available over GA4GH standard interfaces.

Assembled SARS-CoV-2 genomes are periodically retrieved from VirusSeq and lineages are assigned using pangolin (Phylogenetic Assignment of Named Global Outbreak LINeages), a tool developed to implement the Pango nomenclature for SARS-CoV-2 lineages. To ensure that lineage assignments are as accurate as possible, the more accurate but slower UShER mode of pangolin is used to assign lineage. Additionally, since pangolin nomenclature and designations are continuously updated as new variants are sequenced and categorized, both pangolin and its underlying databases are updated in sync with new releases. Upon update to pangolin or its databases, all previously assigned lineages are re-assigned using the most up-to-date databases.

In addition to assigning lineage, the pipeline also produces a single-line multifasta and, for each assembly, the set of sites that differs from the SARS-CoV-2 reference genome. The resulting metadata, variant sites, assemblies, and multifasta are processed through an ingestion pipeline and connected to Viral AI where the data is made publicly available for further analysis and interpretation. Following its ingestion into Viral AI, the lineage metadata is retrieved and added to the VirusSeq Data Portal, and remains crucial to the researchers conducting variant surveillance.

## 2.  Supplementary Results

### 2.1  List of Selected Contextual Data Fields Available on the Data Portal

- Study ID
- Specimen Collector Sample ID
- Sample Collected By
- Sequence Submitted By
- Submission Date
- Sample Collection Date
- Sample Collection Date Null Reason
- Lineage Name
- Lineage Analysis Software Name
- Lineage Analysis Software Version
- Lineage Analysis Software Data Version
- Scorpio Call
- Scorpio Version

- Geo_loc_name (Country)
- Geo_loc_name (State/province/territory)
- Organism
- Isolate
- Fasta Header Name
- Purpose Of Sampling
- Purpose Of Sampling Details
- Anatomical Material
- Anatomical Part
- Body Product
- Environmental Material
- Environmental Site
- Collection Device
- Collection Method
- Host (Scientific Name)
- Host Disease
- Host Age
- Host Age Null Reason
- Host Age Unit
- Host Age Bin
- Host Gender
- Purpose Of Sequencing
- Purpose Of Sequencing Details
- Sequencing Instrument
- Sequencing Protocol
- Raw Sequence Data Processing Method
- Dehosting Method
- Consensus Sequence Software Name
- Consensus Sequence Software Version
- Breadth Of Coverage Value
- Depth Of Coverage Value
- Reference Genome Accession
- Bioinformatics Protocol
- Gene Name
- Diagnostic Pcr Ct Value
- Diagnostic Pcr Ct Value Null Reason

*For a complete list of Data Portal policies and available contextual data, view* [https://virusseq-dataportal.ca/policies](https://virusseq-dataportal.ca/policies)

## 3. Consortium and Network Author Information

**Canadian Public Health Laboratory Network (CPHLN) members and staffs having contributed data to the portal**

**Alberta**

*Genomes with prefix ABPHL*

Bu J, Croxen M, Deo A, Dieu P, Dong X, Ferrato C, Gavriliuc S, George R, Getachew F, Gill, K, Ie N, Khadka R, Khan F, Koleva P, Lee L, Li V, Lindsay A, Lloyd C, Lynch T, Ma R,

McCullough, E, Mohon A, Murphy S, Obasuyi O, Pabbaraju K, Presbitero A, Rotich S, Shokoples S, Thayer J, Tipples G, Trevor H, Whitehouse M, Wong A, Yu C, Zelyas N

*Genomes in collaboration with University of Calgary (prefix AB-NNNNNN)*

Gordon P, Lam LG, Pabbaraju K, Wong A, Ma R, Li V, Melin A, Tipples G, Berenger B, Zelyas N, Kellner J, Bernier F, Chui L, Croxen M

**British Columbia**

*BCCDC Public Health Laboratory*
Natalie Prystajecky, Linda Hoang, John R. Tyson, Dan Fornika, Shannon Russell, Kim MacDonald, Kimia Kamelian, Ana Pacagnella, Corrinne Ng, Loretta Janz, Richard Harrigan, Robert Azana, Tara Newman, Jessica Caleta, Sherrie Wang, Janet Fung, Mel Krajden

**Manitoba**

*Cadham Provincial Laboratory collected specimens sequenced at Canada's National Microbiology Lab (NML)\*\**
Paul Van Caeseele, Jared Bullard, David Alexander, Kerry Dust

*Cadham Provincial Laboratory sequenced specimens*
David Alexander, Lori Johnson, Janna Holowick, Joanne Sanders, Adam Hedley, Kerry Dust, Ayo Bolaji, Brooke Cistarelli, Emma Rempel, Paul van Caeseele, Jared Bullard

*Dynacare sequenced specimens*
Hilary Racher, Melissa Desaulnier, Tintu Abraham, Hongbin Li (Impact Genetics, Brampton Ontario)

*NML\*\**
See below for list of NML personnel

**New Brunswick**

*Centre Hospitalier Universitaire Georges L. Dumont*
Beauregard AP., Lyons P., Chacko S., Shaw W., Lacroix J., Allain E., Crapoulet N., Garceau R., Desnoyers G.

*NML\*\**
See below for list of NML personnel

**Newfoundland and Labrador**

*Newfoundland and Labrador Health Services*
*(previously known as "Newfoundland and Labrador - Eastern Health")*
Robert Needle, Yang Yu, Laura Gilbert, George Zahariadis, Geoffrey Woodland, Chris Corkum, Kerri Smith, Phillip Andrews, Matthew Gilmour

*NML\*\**
See below for list of NML personnel

**Nova Scotia**

*QEII Health Sciences Centre\**
Todd Hatchette, Jason LeBlanc, Janice Pettipas, Dan Gaston, Greg McCracken
(\*data tagged post September 14th should include Allana Loder)

*NML\*\**
See below for list of NML personnel

**Ontario**

*Public Health Ontario Laboratory*
Jacob Afelskie, Vanessa G Allen, Rebecca Azzaro, Doonia Bajovic, Philip Banh, Ilse Belgrave, Tom Braukmann, Ashley Carandang, Yao Chen, Claudia Chu, Shawn Clark, Kirby Cronin, Richard de Borja, Rachelle DiTullio, Carla Duncan, Hadi El Roz, Alireza Eshaghi, Nahuel Fittipaldi, Christine Frantz, Dhiraj Gaglani, Nicole Graham, Jonathan B Gubbay, Jennifer L Guthrie, Lawrence Heisler, Daniel Heydari, Mark Horsman, Hadia Hussain, Jason Iraheta, Grace Jeong, Esha Joshi, Sushma Kavikondala, Lisa Kim, Surendra Kumar, Michael Laszloffy, Aimin Li, Michael C.Y. Li, Alex Marchand-Austin, Maria Mariscal, Dean Maxwell, Lisa McTaggart, Fatima Merza, Anupam Mittal, Naadia Mohammed, Esther Nagai, Sandeep Nagra, Shiva Nassori, Paul Nelson, Rima Palencia, John Palmer, Samir N Patel, Stephen Perusini, Nataliya Potapova, Anna Puzinovici, Zarah Rajaei, Christina Rampertab, Himeshi Samarsinghe, Candice Schreiber, Christine Seah, Fatemeh Shaeri, Kapil Shaeri, Kapil Shah, Narisha Shakuralli, Natasha Singh, Karthikeyan Sivaraman, Brenda Stanghini, Ashleigh Sullivan, Vincent Su Bin Cha, Yogi Sundaravadanam, Sarah Teatero, Semra Tibebu, Nobish Varghese, Andre Villegas, Jesse Wang, Matthew Watson, Sichong Xu, Xiao Xu, Kent Young, Sophie Yu, Farhan Yusuf, Sandra Zittermann

*Ontario Institute for Cancer Research*
Jared T. Simpson, Richard de Borja, Paul Krzyzanowski, Bernard Lam, Lawrence Heisler, Michael Laszloffy, Yogi Sundaravadanam, Ilinca Lungu, Lubaina Kothari, Cassandra Bergwerff, Jeremy Johns, Felicia Vincelli, Philip Zuzarte

*McMaster University*
Hooman Derakhshani, Sheridan J.C. Baker, Emily M. Panousis, Ahmed N. Draia, Jalees A. Nasir, Michael G. Surette, Andrew G. McArthur

**Prince Edward Island**

*Queen Elizabeth Hospital*
Xiaofeng Ding, Vanessa Arseneau, Kari-Lyn Young
*NML\*\**
See below for list of NML personnel

**Québec**

*Laboratoire de Santé Publique du Québec, McGill Génome Sciences Centre; CoVSeq Consortium*
Sandrine Moreira, Jiannis Ragoussis, Guillaume Bourque, Éric Fournier, Aurélie Guilbault, Benjamin Delisle, Dihya Baloul, Inès Levade, Sarah Reiling, Hector Galvez, Paul Stretenowich, Alexandre Montpetit, Michel Roger, Judith Fafard

**Saskatchewan**

*Roy Romanow Provincial Laboratory*
Ryan McDonald, Keith MacKenzie, Kara Loos, Stefani Kary, Meredith Faires, Guruprasad Janga, Rachel DePaulo, Laura Klassen, Alanna Senecal, Amanda Lang, Jessica Minion, Roy Romanow Provincial Laboratory - Molecular Diagnostics

*NML\*\**
See below for list of NML personnel

**Canada's National Microbiology Laboratory (NML)\*\***

Anna Majer, Shari Tyson, Grace Seo, Philip Mabon, Elsie Grudeski, Rhiannon Huzarewich, Russell Mandes, Anneliese Landgraff, Jennifer Tanner, Natalie Knox, Morag Graham, Gary Van Domselaar, Nathalie Bastien, Ruimin Gao, Cody Buchanan, Jasmine Frost, Ameet Bharaj, Cole Slater, Nikki Toledo, Laura Hart, Yan Li, Timothy Booth, Catherine Yoshida, Genevieve Labbe, Adina Bujold, Kara Loos, Jennifer Beirnes, Michael Przybytkowski, Patrick Bastedo, Debra Sorensen, Andrea Tyler, Ana Duggan, Darian Hole, Madison Chapel, Kristen Biggar, Emily Haidl, Chanchal Yadav, Jeff Tuff, Connor Chato, Katherine Eaton, Kirsten Palmier, Molly Pratt, Amber Papineau, Adrian Zetner, Carmen Lìa Murall

Genomics Core Facility at NML
Robotics Support Laboratory at NML

**Academic, Health Network and Research Institutions staffs having contributed data to the Portal**

*Unity Health Toronto*
Ramzi Fattouh, Larissa M. Matukas, Yan Chen, Mark Downing, Trina Otterman, Karel Boissinot, Le Luu

*University Health Network/Mount Sinai Hospital Department of Microbiology*
Marie-Ming Aynaud, Javier Hernandez, Seda Barutcu, Kin Chan, Jessica Bourke, Marc Mazzulli, Tony Mazzulli, Laurence Pelletier, Jeff Wrana, Aimee Paterson, Angel Liu, Allison McGeer

*Kingston Health Sciences Centre and Queen's University*
Prameet M. Sheth, Calvin Sjaarda, Robert Colautti, Katya Douchant

*University of British Columbia*
John R. Tyson, Gabrielle Jayme, Karen Jones, Terrance P. Snutch

*Toronto Invasive Bacterial Diseases Network; Sunnybrook Health Sciences*
Allison McGeer, Patryk Aftanas, Angel Li, Kuganya Nirmalarajah, Emily Panousis, Ahmed Draia, Jalees Nasir, David Richardson, Michael Surette, Samira Mubareka, Andrew G. McArthur

*Eastern Ontario Regional Laboratory Association*
Leanne Mortimer, Hooman Derakhshani, Emily Panousis, Ahmed Draia, Jalees Nasir, Robert Slinger, Andrew G. McArthur

*Western University, CoVizu$_{vs}$ Dev Team*
Art Poon, Gopi Gugan, Bonnie Lu, Roux-Cil Ferreira, Molly Liu, Laura Muñoz Baena, Kaitlyn Wade, Navaneeth Mohan, Sandeep Thokala, Abayomi Olabode

**CanCOGeN VirusSeq committees and working groups' members**

*CanCOGeN VirusSeq Implementation Committee*
Terrance Snutch, Fiona Brinkman, Marceline Côté, William Hsiao, Gordon Jolly, Yann Joly, Sharmistha Mishra, Sandrine Moreira, Samira Mubareka, Jared Simpson, Megan Smallwood, Gary Van Domselaar

*CanCOGeN Capacity Building Working Group*
Gary Van Domselaar, Matthew Croxen, Natalie Knox, Celine Nadon, Jennifer Tanner

*CanCOGeN Data Analytics Working Group*
Gary Van Domselaar, Fiona Brinkman, Zohaib Anwar, Robert Beiko, Matieu Bourgey, Guillaume Bourque, Richard de Borja, Ahmed Draia, Jun Duan, Marc Fiume, Dan Fornika, Eric Fournier, Erin Gill, Paul Gordon, Emma Griffiths, Jose Hector Galvez Lopez, Darian Hole, William Hsiao, Jeffrey Joy, Kimia Kamelian, Natalie Knox, Philip Mabon, Finlay Maguire, Tom Matthews, Andrew McArthur, Samir Mechai, Sandrine Moreira, Art Poon, Amos Raphenya, Claire Sevenhuysen, Jared Simpson, Jennifer Tanner, Lauren Tindale, John Tyson, Geoff Winsor, Nolan Woods, Matthew Croxen, Carmen Lìa Murall

*CanCOGeN Ethics and Governance Working Group*
Yann Joly, Fiona Brinkman, Erin Gill, William Hsiao, Hanshi Liu, Sandrine Moreira, Gary Van Domselaar, Ma'n Zawati, Sarah Savić-Kallesøe

*CanCOGeN Metadata Working Group*
William Hsiao, David Alexander, Zohaib Anwar, Nathalie Bastien, Tim Booth, Guillaume Bourque, Fiona Brinkman, Hughes Charest, Caroline Colijn, Matthew Croxen, Guillaume Desnoyers, Rejean Dion, Damion Dooley, Ana Duggan, Leah Dupasquier, Kerry Dust, Eleni Galanis, Emma Garlock, Erin Gill, Gurinder Gopal, Tom Graefenhan, Morag Graham, Emma Griffiths, Linda Hoang, Naveed Janjua, Jeffrey Joy, Kimia Kamelian, Lev Kearney, Natalie Knox, Theodore Kuschak, Jason LeBlanc, Yan Li, Anna Majer, Adel Malek, Ryan McDonald, David Moore, Celine Nadon, Samir Patel, Natalie Prystajecky, Anoosha Sehar, Claire Sevenhuysen, Garrett Sorensen, Laura Steven, Lori Strudwick, Marsha Taylor, Shane Thiessen, Gary Van Domselaar, Adrian Zetner

*CanCOGeN Research Collaborations Working Group*
Fiona Brinkman, Zohaib Anwar, Marceline Côté, Marc Fiume, Laura Gilbert, Erin Gill, Paul Gordon, Yann Joly, Sandrine Moreira, Samira Mubareka, Natalie Prystajecky, Jennifer Tanner, Gary Van Domselaar, Phot Zahariadis,

*CanCOGeN Sequencing Working Group*
Ioannis Ragoussis, Terrance Snutch, Patryk Aftanas, Matthew Croxen, Hooman Derakhshani, Nahuel Fittipaldi, Morag Graham, Andrew McArthur, Sandrine Moreira, Samira Mubareka, Natalie Prystajecky, Ioannis Ragoussis, Jared Simpson, Michael Surette, John Tyson

*CanCOGeN Quality Control Working Group*
Jared Simpson, Mathieu Bourgey, Kodjovi Dodji Mlaga, Nahuel Fittipaldi, Jose Hector Galvez Lopez, Natalie Knox, Genevieve Labbe, Pierre Lyons, Philip Mabon, Finlay Maguire, Anna Majer, Andrew McArthur, Ryan McDonald, Sandrine Moreira, Natalie Prystajecky, Karthikeyan Sivaraman, Kerri Smith, Terrance Snutch, Karthikeyan Sivaraman, Andrea Tyler, John Tyson, Gary Van Domselaar, Matthew Croxen

*Canadian VirusSeq Data Portal (CVDP) Team*
Guillaume Bourque, Lincoln Stein, Christina Yung, Hanshi Liu, Yann Joly, Adrielle Houweling, William Hsiao, Marc Fiume, David Bujold, Erin Gill, Fiona Brinkman, Nithu John, Rosita Bajari, Linda Xiang, Alexandru Lepsa, Jaser Uddin, Justin Richardsson, Leonardo Rivera

Funding for the VirusSeq Data Portal is provided by The Canadian COVID Genomics Network (CanCOGeN), and supported by Genome Canada and Innovation, Science and Economic Development Canada (ISED)